\documentclass[%
 reprint,
 amsmath,amssymb,
 aps,prc
]{revtex4-2}
\usepackage[colorlinks, citecolor=red]{hyperref}
\usepackage{graphicx}
\usepackage{dcolumn}
\usepackage{bm}
\usepackage{nameref}
\usepackage{natbib}
\usepackage[T1]{fontenc}
\usepackage{booktabs, array, mathptmx, float, tabularx, booktabs, lipsum, amsmath,multirow}
\usepackage{siunitx, xcolor}
\usepackage[version=4]{mhchem}
\begin{document}
\preprint{APS/123-QED}
\title{Nonlocality Effect in the Tunneling of Alpha Radioactivity with the Aid of Machine Learning}

\author{ Jinyu Hu$^{1}$ and  Chen Wu$^{1}$ } \affiliation{
\small 1. Xingzhi College, Zhejiang Normal University, Jinhua, 321004, Zhejiang, China}
\begin{abstract}
   Recently, building upon the research findings of E. L. Medeiros \cite{medeiros2022nonlocality}, we have extended the \texorpdfstring{$\alpha$}{}-particle nonlocality effect to the two-potential approach (TPA) \cite{hu2025nonlocality}. This extension demonstrates that the integration of the \texorpdfstring{$\alpha$}{}-particle nonlocality effect into TPA yields relatively favorable results. In the present work, we employ machine learning methods to further optimize the aforementioned approach, specifically utilizing three classical machine learning models: decision tree regression, random forest regression, and XGBRegressor.
   Among these models, both the decision tree regression and XGBRegressor models exhibit the highest degree of agreement with the reference data, whereas the random forest regression model shows inferior performance. In terms of standard deviation, the results derived from the decision tree regression and XGBRegressor models represent improvements of 54.5\% and 53.7\%, respectively, compared to the TPA that does not account for the coordinate-dependent effective mass of \texorpdfstring{$\alpha$}{} particles.
   Furthermore, we extend the decision tree regression and XGBRegressor models to predict the \texorpdfstring{$\alpha$}{}-decay half-lives of 20 even-even nuclei with atomic numbers Z=118 and Z=120. Subsequently, the superheavy nucleus half-life predictions generated by our proposed models are compared with those from two established benchmarks: the improved eight-parameter Deng-Zhang-Royer (DZR)\cite{deng2020improved} model and the new empirical expression (denoted as "New+D") \cite{denisov2024empirical}proposed by V. Yu. Denisov, which explicitly incorporates nuclear deformation effects.
   Overall, the predictions from these models and formulas are generally consistent. Notably, the predictions of the decision tree regression model show a high level of consistency with those of the New+D expression, while the XGBRegressor model exhibits deviations from the other two comparative models.
\end{abstract}

\maketitle

\section{\label{sec:level1}INTRODUCTION}
Since the 20th century, and particularly following the discovery of natural radioactive decay, $\alpha$-decay has remained a long-standing research focus in nuclear physics. With advances in experimental techniques enabling the laboratory synthesis of superheavy nuclei, interest in the systematic investigation of $\alpha$-decay processes has grown significantly. This decay mode was first identified as $\alpha$-radiation emitted by uranium and its compounds; Ernest Rutherford initially characterized it as a process wherein a parent nucleus emits a $^{4}$He nucleus. Subsequently, the underlying mechanism of $\alpha$-decay was elucidated by quantum tunneling theory, independently proposed by Gurney and Condon \cite{gurney1928wave} and by Gamow \cite{gamow1928quantentheorie}.Since then, numerous researchers have explored key nuclear characteristics of $\alpha$-decay, including nuclear shape coexistence, low-lying states, shell closure effects, energy level structures, and ground-state properties. \cite{matsuse1975study,wauters1994fine,kucuk2020role,wang2017competition,xiao2020alpha}. Furthermore, studies of $\alpha$-decay processes serve as a critical tool for the identification of synthesized heavy and superheavy nuclei.
\par{}
Driven by advances in modern detector technology and the development of realistic nuclear potentials, significant progress has been achieved in both the theoretical \cite{buck1992alpha,duarte1998cold,goncalves1993effective,basu2003role,chowdhury2006alpha} and experimental \cite{ma2020short,oganessian2011eleven,oganessian2004experiments,zhang2021new,oganessian2000observation,oganessian2011synthesis} investigations of $\alpha$-decay.
\par{}
On the theoretical side, numerous models and approaches have been proposed to study $\alpha$-decay, including the two-potential approach \cite{gurvitz1987decay}, fission-like model \cite{buck1993half}, density-dependent M3Y effective interaction \cite{samanta2007predictions}, cluster-formation model \cite{ahmed2017alpha}, and generalized liquid drop model \cite{zhang2006alpha}. Complementing these theoretical frameworks, a wealth of empirical formulas-rooted in the classic Geiger-Nuttall (G-N) law and quantum tunneling effect: have also been developed for this decay mode. Examples include the Viola-Seaborg-Sobiczewski (VSS) formula \cite{viola1966nuclear}, Horoi formula, Universal decay law \cite{qi2009universal}, Royer formula \cite{royer2000alpha}, modified YQZR (MYQZR) formula, Deng-Zhang-Royer (DUR) formula \cite{deng2020improved}, and the new empirical expression accounting for nuclear deformation (denoted as "New+D") \cite{denisov2024empirical}.
\par{}
In terms of experimental progress \cite{hofmann2000discovery,hamilton2013search,nazarewicz2018limits,giuliani2019colloquium}, the synthesis of superheavy elements in laboratories such as GSI, RIKEN, and JINR primarily relies on cold fusion reactions between
$^{208}$Pb or $^{209}$Bi targets and nuclear beams with mass number A>50. As an extension of the periodic table, the superheavy element with atomic number Z=118 was successfully synthesized via $^{48}$Ca-induced hot fusion reactions, using the actinide element californium (Cf) as the target material \cite{oganessian2011synthesis}.
\par{}
Recently, machine learning has attracted significant attention from researchers in physics due to its capability to address complex problems in nonlinear systems. For example, in nuclear physics, it has been applied to predict nuclear masses, $\beta$-decay half-lives, $\alpha$-decay half-lives, charge radii, and neutron drip line. Specifically, the Bayesian neural network method has been successfully applied to predict charge radii, $\beta$-decay half-life, and fission. As a result, it has become a research approach that attracts considerable attention in nuclear physics. In addition, numerous machine learning methods-such as artificial neural networks and convolutional neural networks-are actively utilized in nuclear physics research. More importantly, support vector machines have achieved significant breakthroughs in this field in recent years. Specifically, when Amir Jalili \cite{jalili2024decay} et. al used support vector machines to conduct a detailed calculation of $\alpha$-decay half-lives, with the root mean square errors of 0.427, which is significantly lower than that of other models for calculating $\alpha$-decay half-lives. It is precisely due to the diverse successes of machine learning in the field of nuclear physics that we are motivated to employ this approach to assist in predicting $\alpha$-decay of even-even nuclei.
\par{}
In this paper, we systematically study the nonlocality effect in $\alpha$-decay of even-even nuclei by using machine learning, specifically using decision tree regression, random forest regression, and XGBRegressor models, to optimize the coordinate-dependent effective mass of mass parameters for alpha particles. What's more, based on the research by E. L. Medeiros \cite{medeiros2022nonlocality} et al, we use machine learning models to optimize the coordinate-dependent effective mass parameters for each individual nucleus. It is found that the results obtained by the decision tree regression model and the XGBRegressor model have improved by 54.5\% and 53.7\%, respectively, compared with the two-potential approach without the coordinate-dependent effective mass for alpha particles in terms of the standard deviation. As an application, we further extend the present model to predict the $\alpha$ decay half-lives of 20 even-even nuclei with $Z = 118$ and $Z=120$, compared to DUR and New+D. Finally, we find the predictions of the present model and New+D are highly consistent.
\par{}
The remainder of this article is organized as follows. Section \ref{sec:level2} briefly introduces the theoretical framework. Detailed numerical results and discussion are given in Section \ref{sec:level3}. Finally, a summary is provided in Section \ref{sec:level4}.
\begin{table*}[ht]
    \renewcommand{\arraystretch}{1}
    \setlength{\tabcolsep}{0.4cm}
    \centering
    \caption{Validation results of different machine-learning models. }
    \begin{ruledtabular}
    \scalebox{1}{
    \begin{tabular}{cccc}
        Nucleus& RF & DT&XG\\
        \colrule
        MSE  & 0.76 &0.86 &1.02 \\
        $R^{2}$  &0.40 &0.33 &0.20 \\
    \end{tabular}
    }
    \label{tab1}
    \end{ruledtabular}
\end{table*}
\begin{table*}[ht]
    \renewcommand{\arraystretch}{1}
    \setlength{\tabcolsep}{0.4cm}
    \centering
    \caption{The standard deviations between the experimental $\alpha$ decay half-lives and calculated ones with improved TPA by considering nonlocality effect in alpha decay using RF, DT and XG to optimize the mass parameter $\rho_{S}$. }
    \begin{ruledtabular}
    \scalebox{1}{
    \begin{tabular}{ccccc}
        Nucleus& RF & DT&XG&Without Nonlocality effect\\
        \colrule
        $\sigma$  &0.306 &0.264 &0.259&0.573 \\
    \end{tabular}
    }
    \label{tab2}
    \end{ruledtabular}
\end{table*}
\begin{table*}[ht]
    \renewcommand{\arraystretch}{1}
    \setlength{\tabcolsep}{0.4cm}
    \centering
    \caption{The $\rho_{S}$ adjustments for  four nuclei such as $_{54}^{108}$Xe,  $_{56}^{114}$Ba,  $_{80}^{184}$Hg  and  $_{114}^{290} $Fl. The $\rho_{S}$ parameter is adjusted to minimize the differences between the experimental and calculated half-lives for corresponding nuclei. lg$_{1/2}^{\mathrm{DT}}$, lg$_{1/2}^{\mathrm{TPA}}$ and lg$_{1/2}^{\mathrm{exp}}$  are the logarithms of the corresponding calculated half-life, in s.}
    \begin{ruledtabular}
    \scalebox{1}{
    \begin{tabular}{ccccc}
        Nucleus& $\rho_{S}$ & $\mathrm{lg}_{1/2}^{\mathrm{DT}}$&$\mathrm{lg}_{1/2}^{\mathrm{TPA}}$&$\mathrm{lg}_{1/2}^{\mathrm{exp}}$\\
        \colrule
        $^{108}54$ &0.347 & -4.136 &-4.024 &-4.143 \\
        $^{114}56$ & 2.585 & 1.696 &2.412 &1.694\\
       $^{184}80$ & -0.347 & 3.401 & 3.360 &3.442\\
       $^{290}114$ &-2.182 & 1.907 & 0.486 &1.903\\
    \end{tabular}
    }
    \label{tab3}
    \end{ruledtabular}
\end{table*}
\begin{table*}[ht]
    \renewcommand{\arraystretch}{1}
    \setlength{\tabcolsep}{0.4cm}
    \centering
    \caption{ Predicted $\alpha$ decay half-lives in logarithmic form 20 even-even nuclei with $Z = 118$ and $Z = 120$ using our improved model, DUR, and New+D. The $\alpha$ decay energies are predicted using the WS4+ model.}
    \begin{ruledtabular}
    \scalebox{1}{
    \begin{tabular}{cccccc}
        Nucleus&$Q_{\alpha}^{\mathrm{WS4+}}$&$\mathrm{lg}_{1/2}^{\mathrm{XG}}$&$\mathrm{lg}_{1/2}^{\mathrm{DT}}$& $\mathrm{lg}_{1/2}^{\mathrm{DUR}}$ &$\mathrm{lg}_{1/2}^{\mathrm{New+D}}$\\
        \colrule
        $^{288}118$ &12.587 & -4.304 &-4.307  &-4.826 &-4.234\\
        $^{290}118$ & 12.572 & -4.324 &-4.327 &-4.830 &-4.179 \\
       $^{292}118$ & 12.212 & -3.703	& -3.650	&-4.139 &	-3.570\\
       $^{294}118$ &12.171 & -3.598& -3.567 &-4.081 &-3.413\\
       $^{296}118$ &11.726 & -3.115& -2.564 &-3.095	&-2.405\\
       $^{298}118$ &12.158 & -3.642& -3.611 &-4.117 &-3.475\\
       $^{300}118$ &11.932 & -3.137& -3.136&-3.640 &-2.895 \\
     $^{302}118$ &12.018 & -3.329 & -3.328  &-3.825	&-3.080\\
       $^{304}118$ &13.101 & -5.699 &-5.634 &-6.168 &-5.514\\
       $^{306}118$ &12.459 & -4.310 & -4.390 &-4.871 &-4.278\\
        $^{290}120$ &13.676 & -6.276&	-5.897 &-6.429	&-5.702 \\
       $^{292}120$ & 13.441 & -5.593& -5.506 &-6.012	&-5.306  \\
       $^{294}120$ & 13.215 & -5.160 & -5.094 &-5.601	&-4.978\\
        $^{296}120$ &13.316 & -5.429	& -5.363	&-5.884&-5.286\\
        $^{298}120$ &12.981 & -4.461	& -4.542&-5.057 &-4.325\\
        $^{300}120$ &13.294 & -5.467	 & -5.400&-5.907	& -5.147\\
       $^{302}120$ &12.866& -4.507 & -4.588&-5.071&-4.252 \\
     $^{304}120$ &12.740 & -4.396& 	-4.364&-4.840&-3.889\\
       $^{306}120$ &13.765 & -6.691& -6.378&-6.903	&-6.167\\
       $^{308}120$ &13.501& -4.819&-4.900	&-5.334	&-4.569 \\
    \end{tabular}
    }
    \label{tab4}
    \end{ruledtabular}
\end{table*}
\begin{table*}[ht]
    \renewcommand{\arraystretch}{1}
    \setlength{\tabcolsep}{0.4cm}
    \centering
    \caption{Comparsion the $\sigma$ in the range 52 $\le$ Z $\le 118$. These theoretical calculations are taken from the refs \cite{jalili2024decay,yang2026alpha}.}
    \begin{ruledtabular}
    \scalebox{1}{
    \begin{tabular}{ccccc}
        Nucleus& Amir Jalili et al & Haitao Yang et al &This work(XG)&This work(DT)\\
        \colrule
        $\sigma$  &0.44 &0.306 &0.259&0.264\\
    \end{tabular}
    }
    \label{tab5}
    \end{ruledtabular}
\end{table*}
\section{\label{sec:level2}THEORETICAL FRAMEWORK }
 \subsection{TPA framework}\label{A}
The half-life $T_{1/2}$ for $\alpha$ decay could be determined by $\alpha$ decay width $\Gamma$. In this work, the expression of the $\alpha$ decay half-life is as follow:
\begin{equation}\label{eq1}
T_{1/2}= \frac{\mathrm{\hbar ln 2}}{\Gamma},
\end{equation}
where $\hbar$ is the reduced Planck constant.
In the framework of TPA, the Gamow formula is improved by adding a preexponential factor. The $\alpha$ decay width $\Gamma$ depends on the $\alpha$ preformation factor $P_{\alpha}$, the normalized factor $F$ and the penetration probability $P$, which can be written as:
\begin{equation}\label{eq2}
\Gamma= \frac{\hbar^{2}P_{\alpha} F P}{4\mu},
\end{equation}
where $\mu$ is the reduced mass of the $\alpha$-daughter nucleus system.
The normalized factor $F$, which is given by the integration over the internal region, can be approximated as
\begin{equation}\label{eq3}
F= \frac{1}{\int_{r_{1}}^{r_{2}}\frac{1}{2k(r)}dr },
\end{equation}
where $r$ is the mass center distance between the preformed $\alpha$ particle and daughter nucleus with $k(r) = \sqrt{(\frac{2\mu}{\hbar^{2}}\left | Q_{\alpha} - V(r) \right | )}$ is the wave number of the $\alpha$ particle. $\mu$ denotes the reduced mass of the $\alpha$ particle and daughter nucleus in the center of mass coordinate. $V(r)$ and $Q$ represent the $\alpha$-core potential and $\alpha$ decay energy, respectively. The penetration probability $P$ is calculated by WKB approximation. It can be expressed as:
\begin{equation}\label{eq4}
P=\mathrm{exp} \left [  -2 \int_{r_{2}}^{r_{3}}k(r)dr  \right ],
\end{equation}
where $r_{1}$, $r_{2}$, and $r_{3}$ are the classical turning point, which must satisfy the condition $V(r_{1}) = V(r_{2}) = V(r_{3}) = Q_{\alpha}$. In the inner region ($r_{1}$ < $r$ < $r_{2}$), the nuclear potential commands the state of the preformed $\alpha$ particle. In the outer region ($r_{2}$ < $r$ < $r_{3}$), the electromagnetic interaction has an important role.
\par{}
In 2013, Ahmed et al proposes the cluster formation model to calculate the preformation probability of even-even heavy nuclei \cite{ahmed2013alpha,ahmed2013clusterization}. And, they also extended this model to calculate the odd-A and odd-odd nuclei \cite{ahmed2017alpha}. In this paper, the preformation probability $P_{\alpha}$  required in the decay width $\Gamma$ can be written as:
\begin{equation}\label{eq5}
P_{\alpha}=\frac{E_{f\alpha}}{E}.
\end{equation}
Where $E_{\alpha}$ denotes $\alpha$ cluster-formation energy and $E$ is total energy. For even-even nuclei they can be expressed as
\begin{equation}\label{eq6}
\begin{aligned}
E_{f\alpha} = & 3B(A,Z)+B(A-4,Z-2)   \\&
                   -2B(A-1,Z-1)-2B(A-1,Z),
\end{aligned}
\end{equation}
\begin{equation}\label{eq7}
E=B(A,Z)-B(A-4,Z-2).
\end{equation}
where $B(A,Z)$ is the binging energy of the nucleus with the mass number $A$ and proton number $Z$.
\par{}
The $\alpha$-core potential between the preformed $\alpha$ particle and the daughter nucleus is composed of the nuclear potential $V_{N}(r)$, Coulomb potential $V_{C}(r)$, and centrifugal potential $V_{l}(r)$, which can be written as :
\begin{equation}\label{eq8}
V(r)=V_{N}(r)+V_{C}(r)+V_{l}(r).
\end{equation}
\par{}
In this work, the nuclear potential is described by the type of cosh parametrized form, obtained by analyzing experimental data of $\alpha$ decay. It can be written as:
\begin{equation}\label{eq9}
V_{N}(r)=-V_{0}\frac{1+\mathrm{cosh(R/a)}}{\mathrm{cosh(r/a)}+\mathrm{cosh(R/a)}},
\end{equation}
where $V_{0}$ and $a$ are parameters of the depth and diffuseness of the nuclear potential, respectively. In this paper, the parameter value were obtained with $a = 0.5958$ fm and $V_{0} = 192.42 +31.059 \frac{N-Z}{A}$ MeV \cite{sun2016systematic}, where $N$, $Z$, and $A$ are the neutron, proton, and mass number of daughter nucleus, respectively. $V_{C}(r)$ is the Coulomb potential, which is based on the assumption of the uniformly charged sphere and can be written as
\begin{equation}\label{eq10}
V_{C}(r)  =\left\{\begin{matrix}
 \frac{Z_{d}Z_{\alpha}e^{2}}{2R}\left [  3 - \frac{r^{2}}{R^{2}}\right ] & r\le R \\
  \frac{Z_{d}Z_{\alpha}e^{2}}{r},& r > R,
\end{matrix}\right.
\end{equation}
where $Z_{d}$ and $Z_{\alpha}$ are the charge number of the daughter nucleus and $\alpha$-particle, respectively. The sharp radius of interaction  $R$ is written as:
\begin{equation}\label{eq11}
R=1.28 A^{1/3} - 0.76+ 0.8A^{-1/3}.
\end{equation}
\par{}
For unfavored $\alpha$ decay ($l \ne 0$ decays, where $l$ is the angular momentum obtained by the emitted $\alpha$ particle), the centrifugal potential generated by the nonzero angular momentum can be given as
\begin{equation}\label{eq12}
V_{l}(r)=\frac{\hbar^{2}(l+\frac{1}{2})^{2}}{2\mu r^{2}}.
\end{equation}
\par{}
For $l(l+1)$ $\to $ $(l+\frac{1}{2})^{2}$ is the Langer modified form, because it is an important correction for one-dimensional problems \cite{morehead1995asymptotics}. The minimum angular momentum  $l_{min}$ taken away  by the emitted $\alpha$-particle is selected in based on the conservation laws of spin-parity, which is written as
 by \cite{sun2017systematic}
\begin{equation}\label{eq13}
l_{min}= \left\{\begin{matrix}
 \Delta_{j}, &for  &even  & \Delta_{j} &and  &\pi_{p}=\pi_{d}, \\
  \Delta_{j+1},& for &even  & \Delta_{j} &and  & \pi_{p}\ne \pi_{d},\\
  \Delta_{j},&  for&  odd& \Delta_{j} & and & \pi_{p}=\pi_{d},\\
  \Delta_{j+1},&  for&  odd& \Delta_{j} & and &\pi_{p}\ne \pi_{d},
\end{matrix}\right.
\end{equation}
where $\Delta_{j} = \left | j_{p} - j_{d} \right |$, $j_{p}$, $\pi_{p}$, $j_{d}$, $\pi_{d}$ denote the spin and parity values of parent and daughter nuclei, respectively. Details on the conservation laws of spin-parity are available in the Ref \cite{sun2017systematic}.
 \subsection{Nonlocality Effect}\label{B2}
\par{}
From the research on E. L. Medeiros \cite{medeiros2022nonlocality}, it can be known that the effective mass of the $\alpha$ particle can be defined as:
\begin{equation}\label{eq14}
\mu=\frac{m^{*}M}{m^{*}+M},
\end{equation}
where $M$ is the nuclear mass of the daughter nucleus. In 2013,based on the gradient of velocity-dependent potential,R.A. Zureikat and M.I. Jaghoub defined the spatially variable mass $m^{*}$:
\begin{equation}\label{eq15}
m^{*}=\frac{m}{1-\rho (r)},
\end{equation}
where $m$ is the free mass of $\alpha$ particle. From the studies of R.A. Zureikat and M.I. Jaghoub, we can learn that $\rho (r)$ is an isotropic function of the radial variable $r$. Meanwhile, the $\rho (r)$ represents the change in the mass of the incident particle.
The $\rho(r)$ function is defined in \cite{jaghoub2011novel, zureikat2013surface,alameer2021nucleon}:
\begin{equation}\label{eq16}
\rho(r)=\rho_{S}a_{s}\frac{\mathrm{d}}{\mathrm{d}r}\left [ 1 +\mathrm{exp}(\frac{r - R_{S}}{a_{S}}) \right ]^{-1},
\end{equation}
Based on the research on E. L. Medeiros, we can learn that  the $R_{S}$ parameter is defined as $R_{S} = R + \Delta R$ and $a_{S}$ is related to the width of this function. In this work , We take the values $\Delta R = 3.44$ ($\Delta R = 2 R_{\alpha}$) fm.
\begin{figure}[ht]
    \centering
    \includegraphics[width=0.5\textwidth]{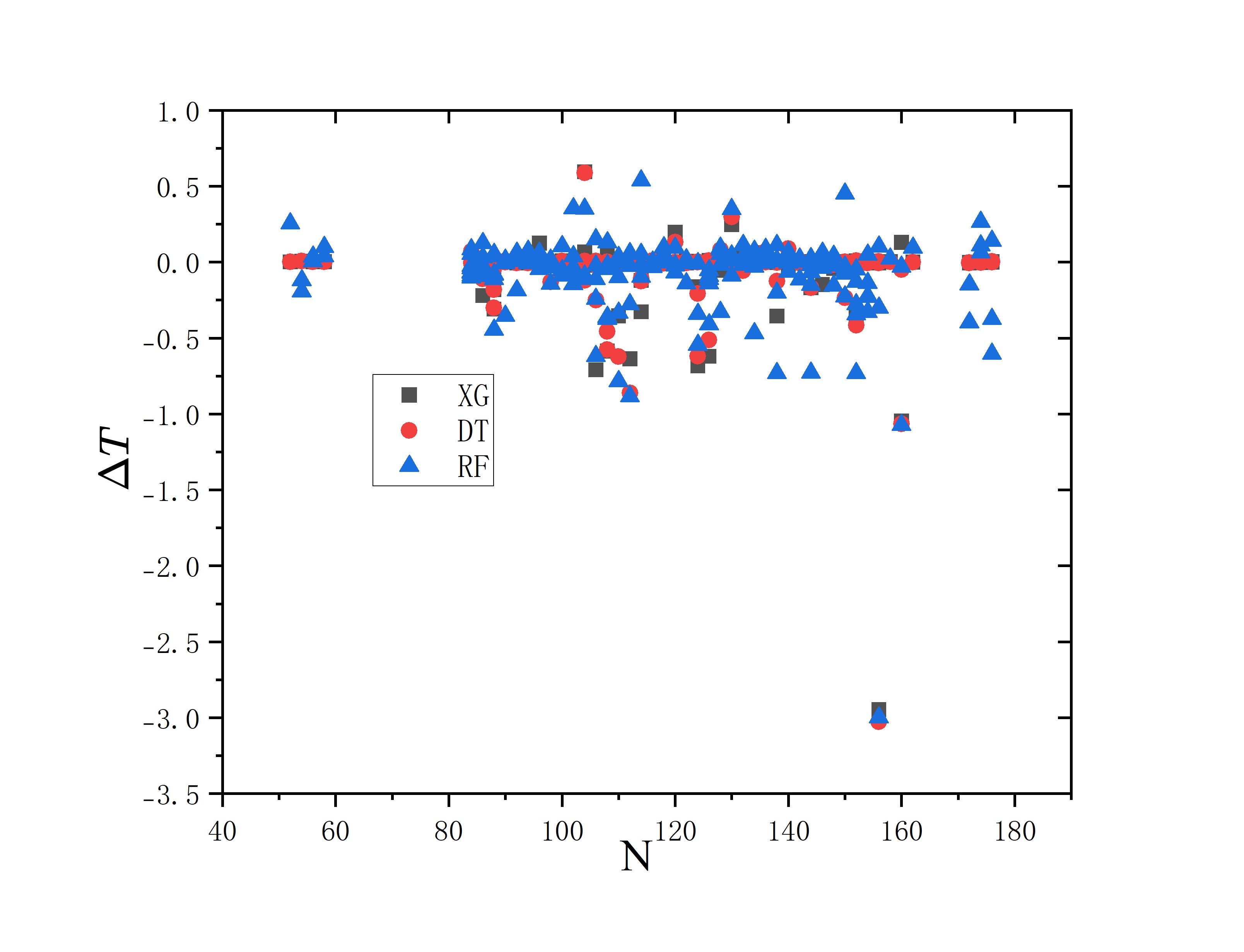}
    \caption{The difference in logarithmic form of $\alpha$ decay half-lives between calculated data and the experimental. The abscissa is neutron number $N$ and the ordinate is the value of  $\mathrm{log_{10}}(T_{1/2}^{\mathrm{cal}}/T_{1/2}^{\mathrm{exp}})$. The red dots, black squares, and blue triangles represent the theoretical calculation results obtained by using the TPA that takes into account the mass parameters of the nonlocality effect optimized by decision tree regression model, XGBRegressor model, and random forest regression model, respectively.}
    \label{imag1}
\end{figure}
\begin{figure*}[ht]
    \centering
    \includegraphics[width=\textwidth]{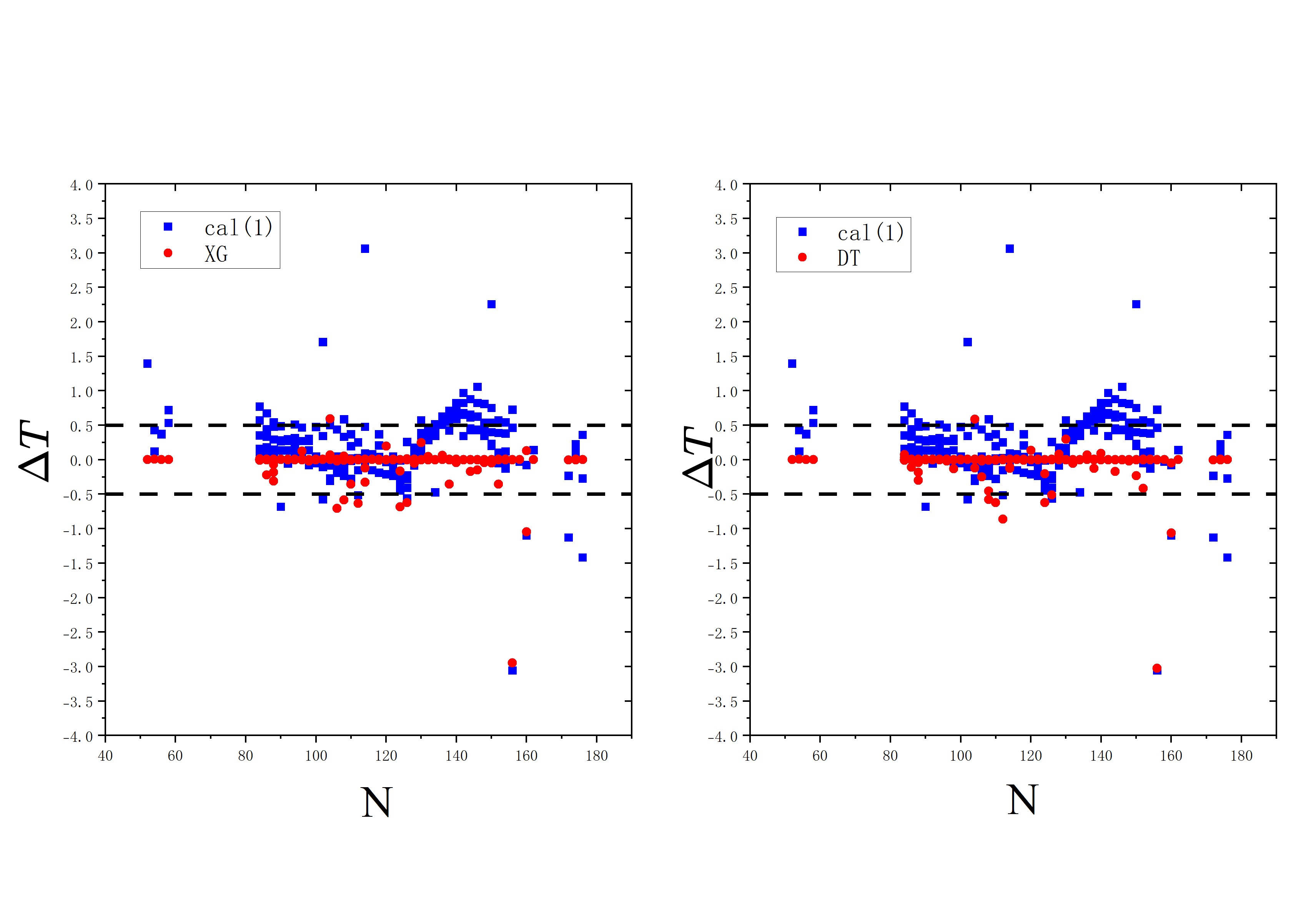}
    \caption{  The difference in logarithmic form of $\alpha$ decay half-lives is shown between the calculated data and the experimental data. The abscissa is neutron number $N$ and the ordinate is the value of  $\mathrm{log_{10}}(T_{1/2}^{\mathrm{cal}}/T_{1/2}^{\mathrm{exp}})$. The red dots in left column and blue squares in left column  represent the theoretical calculation results obtained by using the TPA that takes into account the mass parameters of the nonlocality effect optimized by the XGBoost regression model(XG) and the theoretical calculation results obtained by using the TPA without nonlocality effect, respectively. The red dots in right column and blue squares in left column  represent the theoretical calculation results obtained by using the TPA that takes into account the mass parameters of the nonlocality effect optimized by the Decision Tree regression model(DT) and the theoretical calculation results obtained by using the TPA without nonlocality effect, respectively.}
    \label{imag2}
\end{figure*}
\subsection{Machine learning methods}\label{C3}
\par{}
Machine learning is a powerful technique for addressing a wide range of nonlinear system problems. It leverages large volumes of data to learn patterns and make informed decisions. A core aspect of machine learning involves iterative optimization aimed at minimizing errors-specifically, reducing the discrepancy between predicted outputs and target values during training. This strong learning capability enables machine learning to effectively model and solve complex nonlinear problems. The primary tasks in machine learning can be broadly categorized into regression and classification. In this work, we focus primarily on regression tasks.
\par{}
In this paper, we use supervised learning methods for a regression task. We use three tree-based supervised ML methods Random Forest (RF), Decision Tree (DT), and XGBoost to predict the mass parameter of the nonlocality effect.
\par{}
\textit {Decision Trees:} The Decision Tree (DT) algorithm uses a tree-like data structure to learn hidden data patterns through numerous sets of rules. Decisions are made at the tree nodes based on attribute importance, which is determined by Gini coefficients. It is important to note that the final decision is made at the leaves of the tree.
\par{}
\textit {Random Forest:}
A Random Forest (RF) \cite{breiman2001random} uses bagging techniques, where a random subset of features and data is sampled to train multiple decision trees. It creates a forest of decision trees, from which predictions are made using the generated forest.
\par{}
\textit {XGBoost:} XGBoost (XG) \cite{chen2016xgboost} represents eXtreme Gradient Boosting, a machine learning algorithm within a gradient boosting framework. Built on gradient-boosted decision trees designed for performance, XG achieves speed through parallel tree-boosting techniques. It supports classification and regression tasks, offering more efficient solutions than its counterparts such as DT and RF.

\par{}
Based on the study in Ref. N. Teruya \cite{teruya2016nonlocality}, it is known that the calculated $\alpha$-decay half-life can be brought into agreement with the experimental value by appropriately adjusting the parameter $\rho_{S}$. Accordingly, this parameter- previously determined through manual tuning-is treated as the output variable in the present machine-learning framework. Following Royer's empirical formula, the quantities $\frac{Z}{\sqrt{Q}}$ and $A$ are selected as the input features of the machine-learning model.
In addition, we employ a set of pre-optimized $\rho_{S}$ values obtained from our prior adjustments, for which the deviation between the calculated and experimental $\alpha$-decay half-lives of even-even nuclei in the range $Z$ = 52 to $Z$ = 118 is constrained to be within 0.001. These accurately tuned values $\rho_{S}$ are incorporated into the machine-learning framework as training data, serving as the target outputs for the regression task. The training strategy is based on automatic hyperparameter optimization to efficiently train the entire model. In this work, machine-learning techniques are adopted to automatically tune the hyperparameters of Random Forest (RF), Decision Tree (DT), and eXtreme Gradient Boosting (XG) models using standard automated search strategies implemented in Python~3.7, thereby enhancing the prediction accuracy and reliability of the regression task. Subsequently, the performance of the trained models is quantitatively evaluated using the mean squared error (MSE) and the coefficient of determination ($R^{2}$). Detailed analyses of the model performance and comparative discussions are presented in the Results and Discussion section.

\section{\label{sec:level3}RESULTS AND DISCUSSION}
\par{}
Based on the nonlocality effect in alpha decay, we propose an improved TPA for heavy and superheavy nuclei by using three machine learning methods RF, DT, and XG to optimize the mass parameter for each nucleus. In other words, the values of the adjustable mass parameter are obtained by machine learning methods that predict the experimental fitting values of  $\alpha$ decay half-lives for 196 nuclei ranging from $Z=52$ to $Z=118$ with $N$ $\ge$ 52.
\par{}
In this work, the $\alpha$-decay half-lives of 196 even-even nuclei are calculated using the present theoretical framework. The logarithmic differences between experimental data and the calculated half-lives are presented in Fig. \ref{imag1}. The validation results of the DT, RF, and XG models show that RF performs best in the regression of parameter $\rho_{S}$. Nevertheless, Fig. \ref{imag1} demonstrates that DT and XG yield more accurate predictions for the $\alpha$-decay half-life. Given that the focus of this work is the calculation of the  $\alpha$-decay half-life rather than the optimization of parameter $\rho_{S}$, DT and XG are considered to be more appropriate for this study. To further quantify the role of nonlocality in the $\alpha$-decay process, Fig. \ref{imag2} compares the deviations for three distinct theoretical scenarios: the TPA without nonlocality effects, the TPA incorporating nonlocality effects via the XG parametrization, and the TPA with nonlocality treated using the DT parametrization.
\par{}
The results indicate that the inclusion of nonlocality, as implemented through either the XG or DT formalism, generally yields improved agreement with experimental half-lives over the local TPA baseline. This suggests that nonlocal interactions play a significant role in the description of the $\alpha$-decay process. Notable exceptions are the nuclei $_{108}^{268}$Hs and $_{108}^{264}$Hs for which deviations remain comparatively large. These discrepancies may be attributed to nuclear deformation effects in the Z=108 region, which are not fully captured by the current model.
\par{}
Overall, the deviations between theory and experiment fall predominantly within $\pm 0.5$in logarithmic units, demonstrating that the present model, particularly with the inclusion of nonlocality, provides a sound theoretical description of the $\alpha$-decay half-lives across the isotopic chains considered.
\par{}
The global property i.e. standard deviation $\sigma$ frequently reflects the consistency between the experimental and calculated $\alpha$ decay half-lives, and it can be written as

\begin{equation}\label{eq17}
\sigma =\sqrt{\frac{1}{n}\sum_{i=1}^{n}(\Delta_{i})^{(2)}};\ \Delta_{i}= \mathrm{log_{10}}\frac{(T_{i}^{\mathrm{cal}})}{(T_{i}^{\mathrm{exp}})},
\end{equation}
Here, $T_{i}^{\mathrm{cal}}$ and $T_{i}^{\mathrm{exp}}$ denote the calculated and experimental $\alpha$-decay half-lives of the i-th nucleus, respectively, and $\Delta_{i}$ represents the logarithmic deviation between theory and experiment. The detailed numerical results obtained using the XG, DT, and RF parametrizations are summarized in Table \ref{tab1}.Relative to the original TPA model, the inclusion of nonlocality leads to a significant reduction in the standard deviation: 54.7$\%$ for the XG case, 53.9$\%$ for DT, and 46.7$\%$ for RF. These improvements confirm that the present model with nonlocal corrections reproduces the experimental half-lives consistently well.
\par
\begin{figure*}[b]
    \centering
    \includegraphics[width=1.2\textwidth]{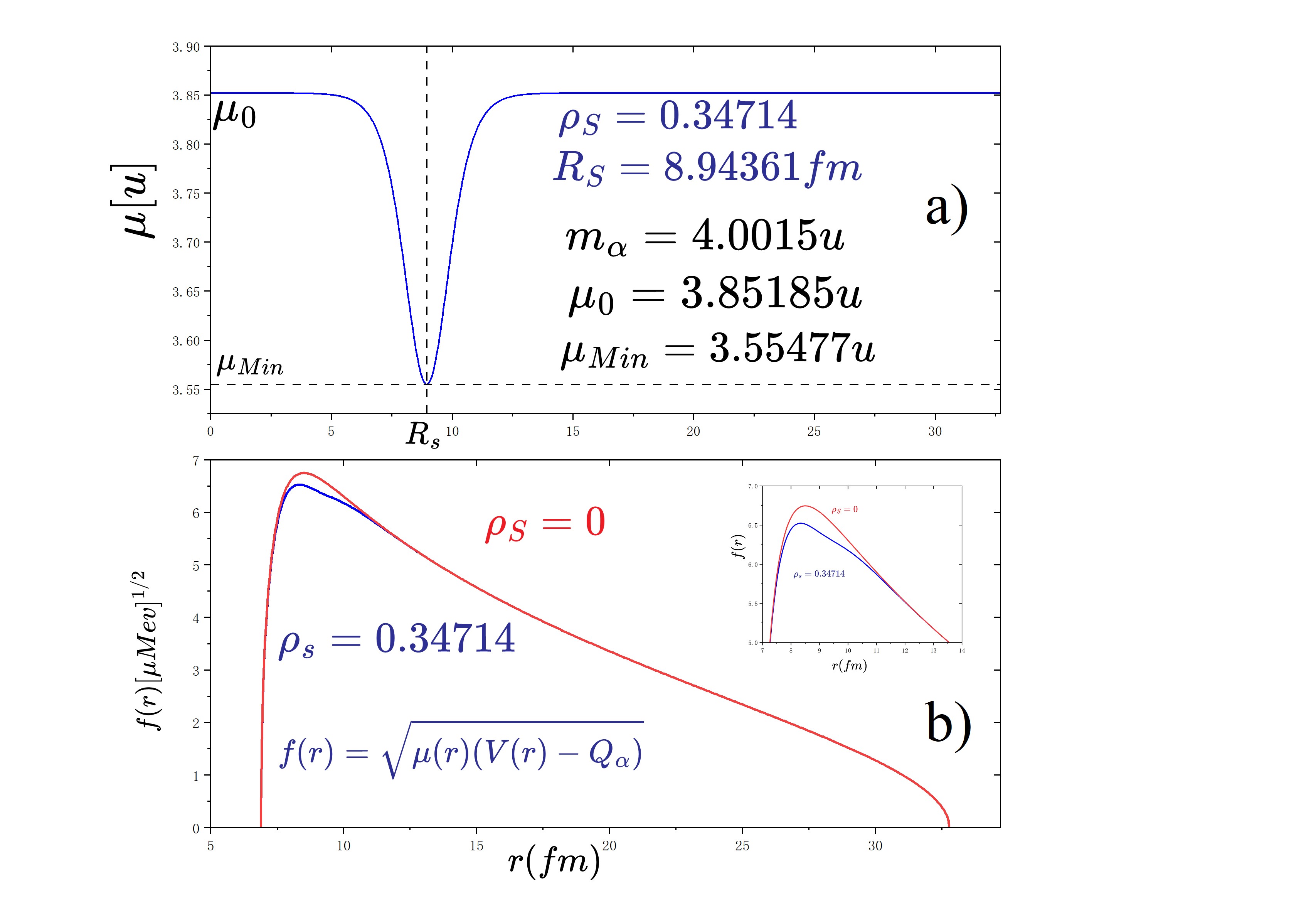}
    \caption{  The contribution of the nonlocal effect on tunneling calculations. Selected example for $\alpha$-decay from $_{54}^{108} $Xe : (a) effective reduced mass $\mu$ considering nonlocality effect with $\rho_{s} = 0.34714$; (b) comparison between the functions $f(r)$ in the integrand of the barrier penetration probability: considering the reduced masses $\mu_{0}$ (red line $\rho_{S} = 0$) and $\mu$ (blue line $\rho_{S} = 0.34714$).}
    \label{imag3}
\end{figure*}

\begin{figure*}[b]
    \centering
    \includegraphics[width=1.2\textwidth]{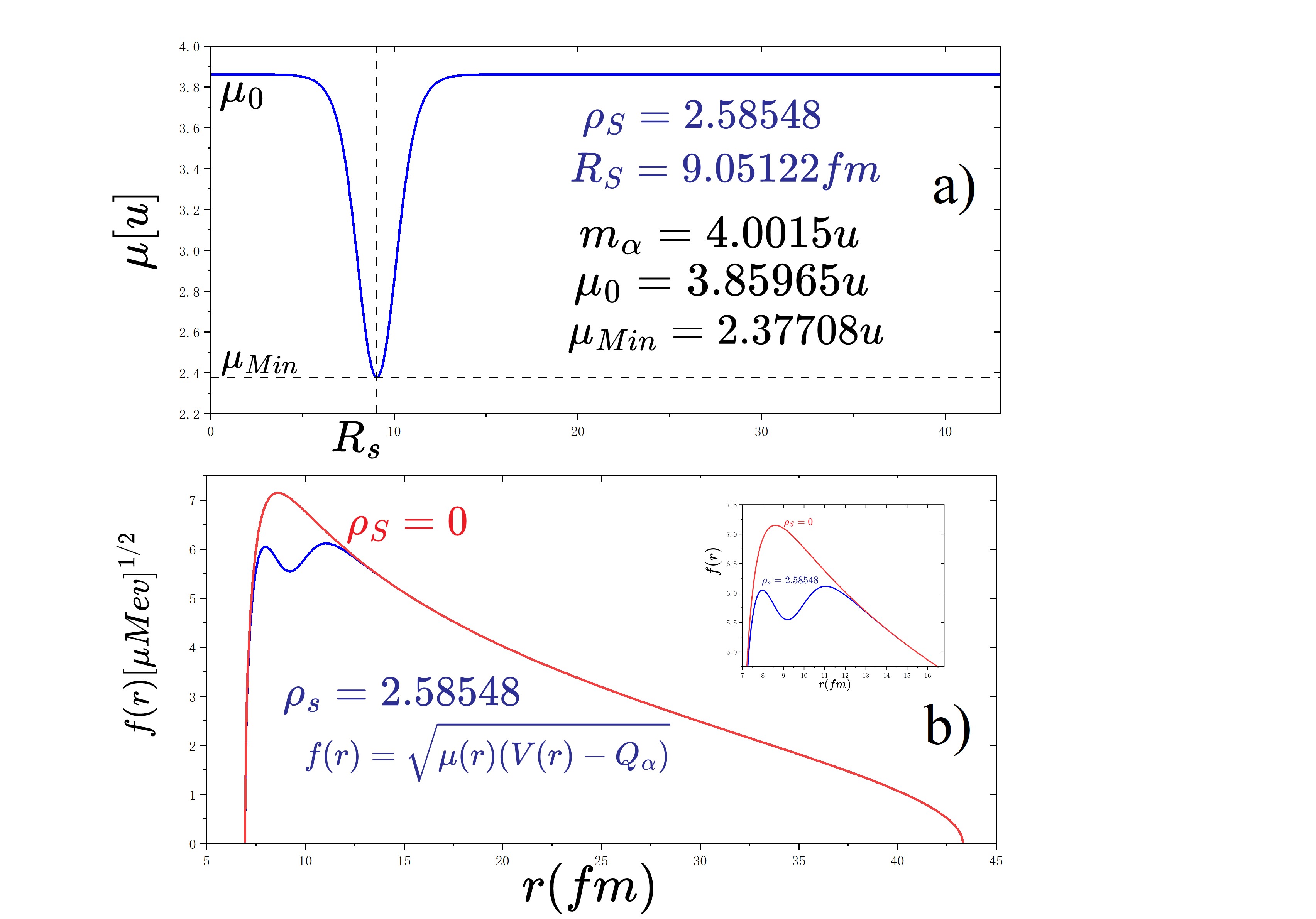}
    \caption{   The contribution of the nonlocal effect on tunneling calculations. Selected example for $\alpha$-decay from $_{56}^{114} $Ba : (a) effective reduced mass $\mu$ considering nonlocality effect with $\rho_{s} = 2.58548$; (b) comparison between the functions $f(r)$ in the integrand of the barrier penetration probability: considering the reduced masses $\mu_{0}$ (red line $\rho_{S} = 0$) and $\mu$ (blue line $\rho_{S} = 2.58548$).}
    \label{imag4}
\end{figure*}

\begin{figure*}[b]
    \centering
    \includegraphics[width=1.2\textwidth]{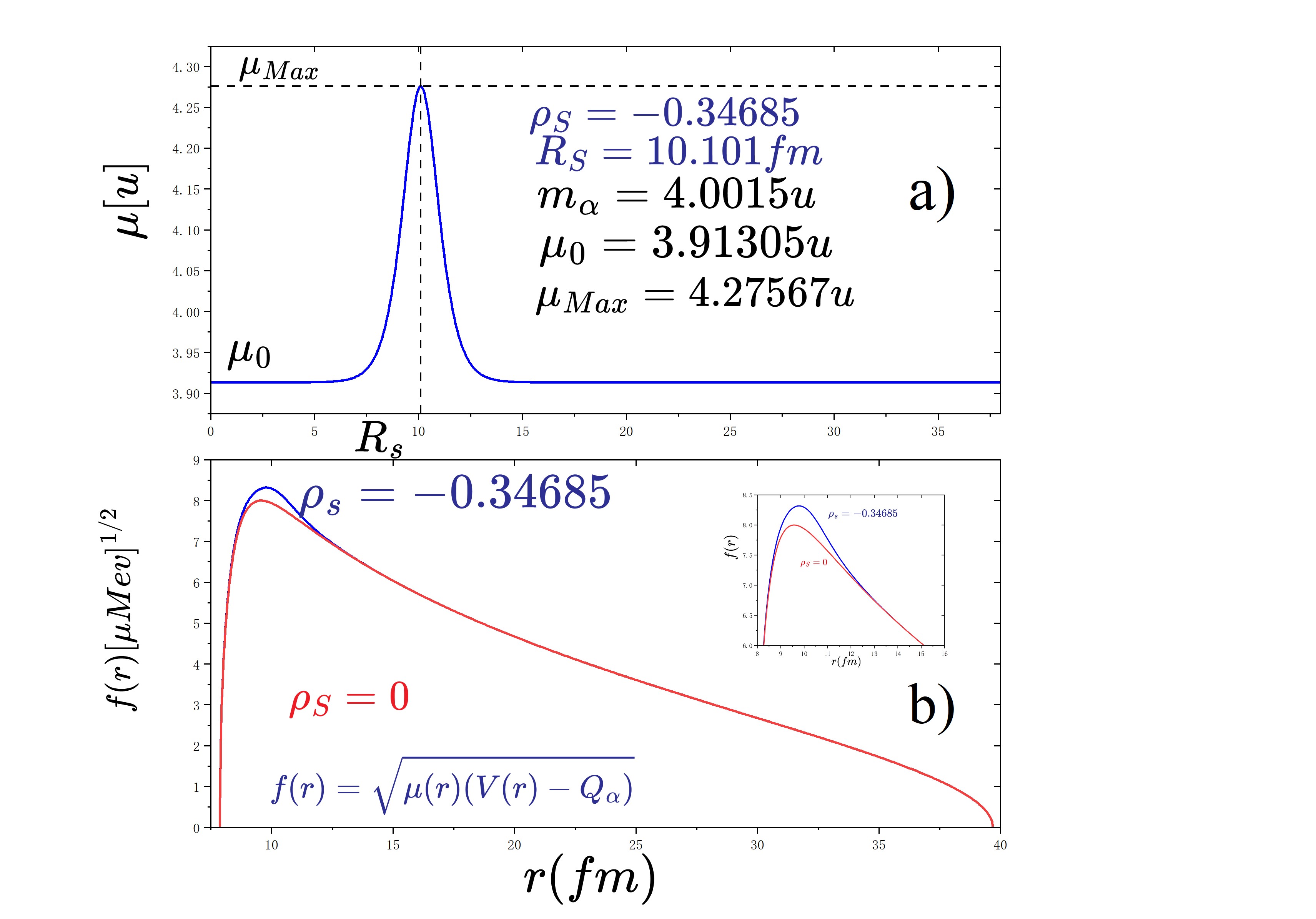}
    \caption{  The contribution of the nonlocal effect on tunneling calculations. Selected example for $\alpha$-decay from $_{80}^{184} $Hg : (a) effective reduced mass $\mu$ considering nonlocality effect with $\rho_{s} = -0.34685$; (b) comparison between the functions $f(r)$ in the integrand of the barrier penetration probability: considering the reduced masses $\mu_{0}$ (red line $\rho_{S} = 0$) and $\mu$ (blue line $\rho_{S} = -0.34685$).}
    \label{imag5}
\end{figure*}

\begin{figure*}[b]
    \centering
    \includegraphics[width=1.2\textwidth]{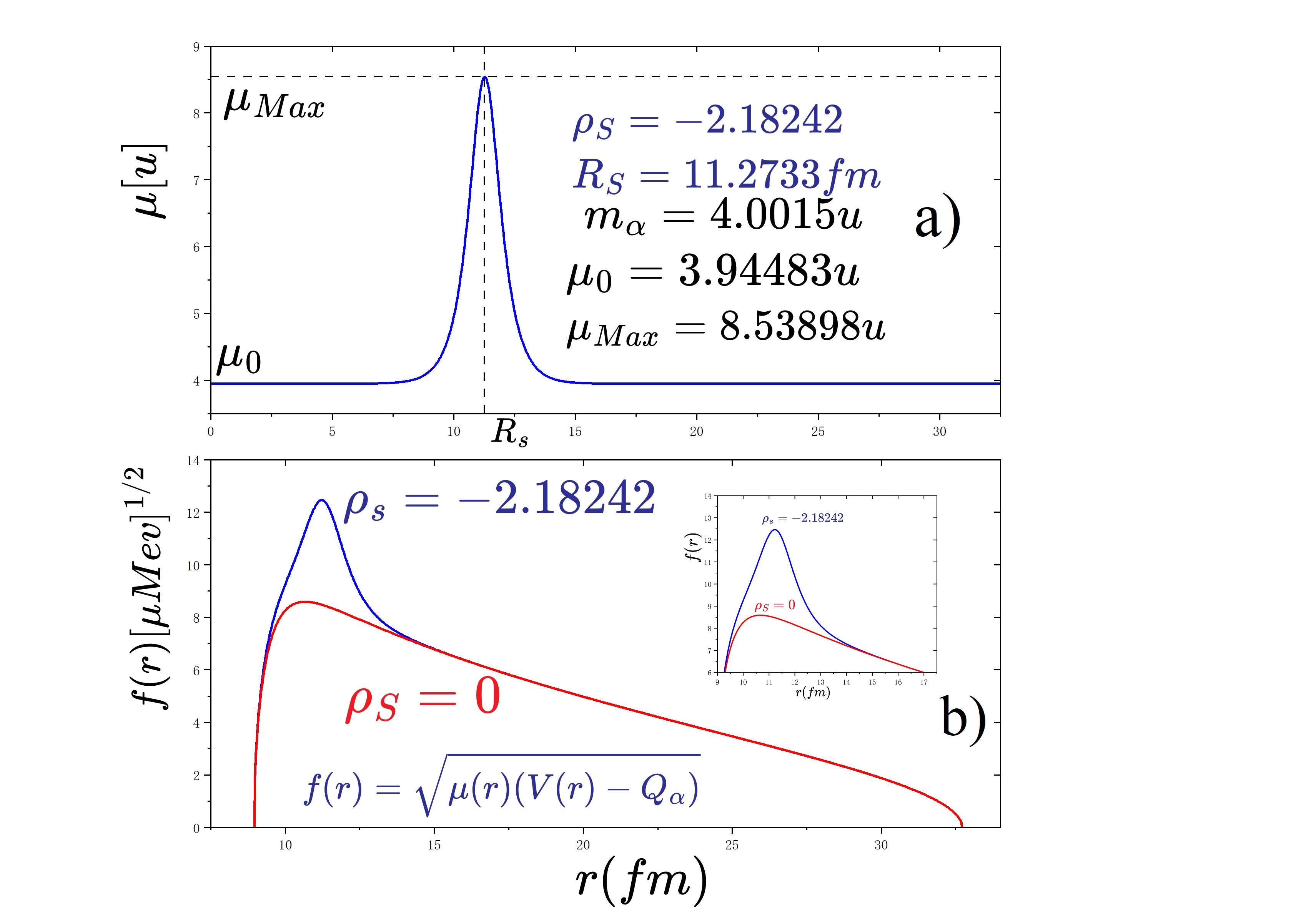}
    \caption{ The contribution of the nonlocal effect on tunneling calculations. Selected example for $\alpha$-decay from $_{114}^{290} $Fl : (a) effective reduced mass $\mu$ considering nonlocality effect with $\rho_{s} = -2.18242$; (b) comparison between the functions $f(r)$ in the integrand of the barrier penetration probability: considering the reduced masses $\mu_{0}$ (red line $\rho_{S} = 0$) and $\mu$ (blue line $\rho_{S} = -2.18242$).}
    \label{imag6}
\end{figure*}

{}
Further insight is provided in Table \ref{tab3}, which illustrates that adjusting the $\rho_{s}$ parameter already brings the calculations closer to experiment, while the inclusion of nonlocality offers an additional improvement. This effect is exemplified in Figs. \ref{imag3}-\ref{imag6} for selected nuclei: For $_{54}^{108} $Xe $(\rho_{s} = 0.34714)$, the deviation is reduced by 93.82$\%$ when nonlocality (DT) is included. For  $_{56}^{114}$Ba $(\rho_{s} = 2.58548)$,the corresponding improvement reaches 99.64$\%$. For $_{80}^{184} $Hg $(\rho_{s} = -0.34685)$,the deviation decreases by 50.67$\%$. In the case of $_{114}^{290} $Fl $(\rho_{s} = -2.18242)$,the agreement becomes nearly perfect, with a 100$\%$ reduction in deviation. These systematic improvements underscore the critical role of nonlocal effects in enhancing the predictive accuracy of the $\alpha$-decay model across a wide range of nuclei.
\par{}
In other words, it is important to note that a consequence of the dynamic effect of the nonlocality of the potential is to produce an increase or decrease of the effective mass of the alpha particle. In the case for $\alpha$-decay of $_{54}^{108}$Xe, $_{56}^{114} $Ba, $_{80}^{184}$Hg, and $_{114}^{290}$Fl  isotopes, this influences the results in relation to the model with the free mass of the particle (see Fig. \ref{imag3}, \ref{imag4}, \ref{imag5} and \ref{imag6}).
\par
\begin{figure*}[b]
    \centering
    \includegraphics[width=1.1\textwidth]{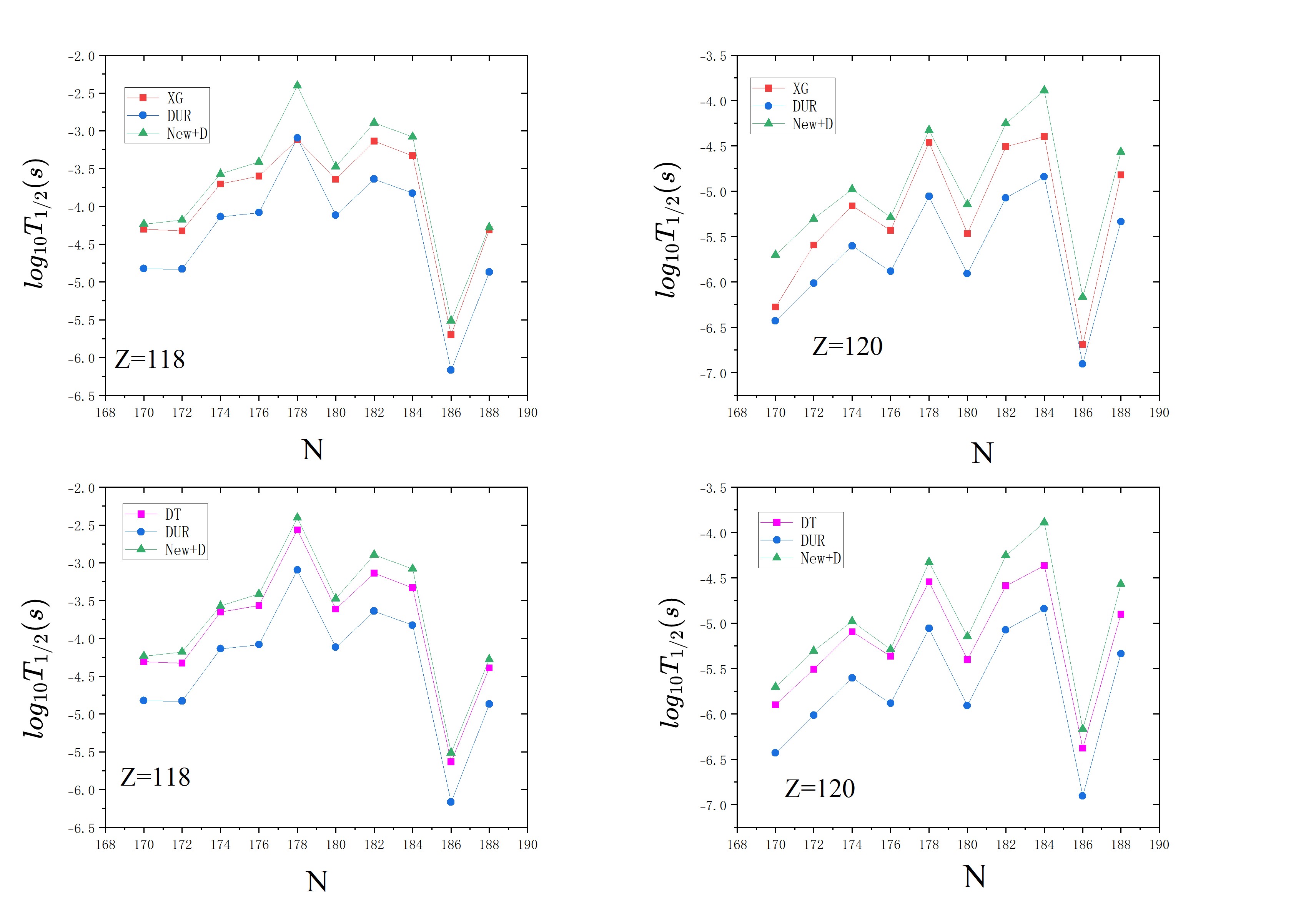}
    \caption{ The predicted values of $\alpha$ decay half-lives for even-even nuclei of $Z = 118$ and $Z = 120$ isotopes are presented. The abscissa is neutron number $N$  and the ordinate is logarithm $\mathrm{log_{10}}T_{1/2}$ of calculated half-life in s. The black squares and red dots indicate $Z = 118$ and $Z = 120$, respectively. The red squares, blue dots and green triangles in the top row represent the theoretical calculation results obtained by using the TPA that takes into account the mass parameters of the nonlocality effect optimized by XGBRegressor model, results obtained by using the DUR model, and results obtained by using the New+D model, respectively. The carmine squares, blue dots and green triangles in the bottom row represent the theoretical calculation results obtained by using the TPA that takes into account the mass parameters of the nonlocality effect optimized by XGBRegressor model, results obtained by using the DUR model, and results obtained by using the New+D model, respectively. The mass tables used in this figure are the WS4 mass tables.}
    \label{imag7}
\end{figure*}

{}
In this section, we employ our improved model to predict the $\alpha$-decay half-lives (in logarithmic form) of 20 even-even nuclei with proton numbers
 $Z=118$and $Z=120$. Accurate prediction of $\alpha$-decay half-lives requires reliable values $Q_{\alpha}$, to which the half-lives are highly sensitive. In this work, we adopt the WS4+RBF model, which has been shown by Ning Wang et al. to reproduce experimental alues of superheavy nuclei most accurately. For comparison, we also incorporate predictions using the well-established DUR and NEW+D models. The corresponding results are summarized in Table \ref{tab4}.
To facilitate visual comparison, the predicted half-lives are plotted in Fig. \ref{imag7} as a function of the parent nucleus neutron number $N$. The results indicate generally consistent predictions across the different models. In particular, the decision tree regression (DTR) and NEW+D models exhibit close agreement, whereas the XGBRegressor (XG) yields relatively divergent predictions.
As seen in Fig. \ref{imag7}, the $\alpha$-decay half-lives generally increase for  $N<178$ peak, and then decrease to a minimum at $N = 180$. A similar trend is observed around $N = 186$. These systematic variations suggest that $N=186$ may correspond to a neutron magic number, while $N=180$ is likely a neutron submagic number in this region.
\par{}
As can be seen from Table \ref{tab5}, our method exhibits certain advantages in calculating the $\alpha$-decay half-lives of even-even nuclei. However, it should be noted that this comparison is relatively rough, due to differences in the datasets used and the fact that our study focuses exclusively on even-even nuclei, in contrast to the works of Amir Jalili et al \cite{jalili2024decay}, Haitao Yang et al \cite{yang2026alpha}. In our approach, machine learning is employed to optimize the parameter ($\rho_{s}$)- free adjustable parameter that governs the effective mass of the $\alpha$ particle during the decay process. This strategy not only improves the accuracy of half-life calculations but also provides a better description of the underlying physical process. By regulating the effective mass, our method allows for the identification of physical factors influencing this quantity, thereby offering deeper insight into the dynamics of $\alpha$-decay. It is important to emphasize that the present work primarily serves as a validation that machine learning can effectively learn the characteristics of the $\rho_{s}$. A more detailed analysis aimed at extracting the key features governing this parameter will be the focus of our subsequent research. A similar methodology has recently been adopted by Yang Haitao et al., who employed machine learning to predict the $\alpha$ preformation probability within the double-folding model (DFM), rather than directly predicting half-lives. This parallel reinforces the value of integrating machine learning with physics-based models to enhance both predictive power and physical interpretability.

\section{\label{sec:level4}SUMMARY AND CONCLUSION}
Based on the nonlocality effect in alpha decay, we use three machine learning methods such as RF, DT, XG to optimize the mass parameter for each nucleus. We proposed an improved TPA for heavy and superheavy nuclei. In this work, we calculate $\alpha$-decay half-lives of 196 nuclei in heavy and superheavy region. The standard deviations are 0.306, 0.264, 0.259 for RF, DT, XG, respectively. Moreover, we also extend the DT and XG model to predict $\alpha$-decay half-lives for 20 even-even nuclei with $Z = 118$ and $Z=120$. In order to make a comparison, the reliable DUR model, the New+D model are used. These predicted results of these model are consistent with each other. And, it is found that the predictions of the decision tree regression model and New+D are highly consistent while the XGBRegressor model deviates from the other two. The predictions imply that 178 and 184 are the probable neutron submagic number and the neutron magic number, respectively. These predicted results can provide useful information for experiments of synthesising new elements and isotopes.
\begin{acknowledgments}
This work was supported by the Zhejiang normal university Doctorial research fund Contract No. ZC302924005.
\end{acknowledgments}


\bibliography{study}
\end{document}